\documentclass{article}
\usepackage{spconf,amsmath,amssymb,graphicx,url}
\usepackage{subfig}
\usepackage{tikz}
\usepackage{comment}
\usepackage{cite}
\usepackage{enumitem}

\newcommand{\Fig}[1]{Fig.~\ref{fig:#1}} 
\newcommand{\Table}[1]{Table~\ref{tab:#1}} 
\newcommand{\Sec}[1]{Section~\ref{sec:#1}} 

\newcommand{\pt}[1]{\left(#1\right)} 

\title{Duration-aware pause insertion using pre-trained language model for multi-speaker text-to-speech}
\name{Dong Yang$^1$, Tomoki Koriyama$^{2,1}$, Yuki Saito$^1$, Takaaki Saeki$^1$, Detai Xin$^1$, Hiroshi Saruwatari$^1$\thanks{Part of this work was supported by JSPS KAKENHI 21K11955.}}
\address{$^1$The University of Tokyo, Japan, $^2$CyberAgent, Inc., Japan.}

\begin{document}
\ninept
\maketitle
\setlength{\tabcolsep}{1mm} 
\setlength{\abovedisplayskip}{5pt} 
\setlength{\belowdisplayskip}{5pt} 
\setlength\floatsep{12pt} 
\setlength\intextsep{12pt} 
\setlength\textfloatsep{12pt} 
\setlength\abovecaptionskip{8pt} 
\setlength{\dbltextfloatsep}{8pt} 
\setlength{\dblfloatsep}{8pt}

\begin{abstract}
Pause insertion, also known as phrase break prediction and phrasing, is an essential part of TTS systems because proper pauses with natural duration significantly enhance the rhythm and intelligibility of synthetic speech. However, conventional phrasing models ignore various speakers' different styles of inserting silent pauses, which can degrade the performance of the model trained on a multi-speaker speech corpus. To this end, we propose more powerful pause insertion frameworks based on a pre-trained language model. Our approach uses bidirectional encoder representations from transformers (BERT) pre-trained on a large-scale text corpus, injecting speaker embeddings to capture various speaker characteristics. We also leverage duration-aware pause insertion for more natural multi-speaker TTS. We develop and evaluate two types of models. The first improves conventional phrasing models on the position prediction of respiratory pauses (RPs), i.e., silent pauses at word transitions without punctuation. It performs speaker-conditioned RP prediction considering contextual information and is used to demonstrate the effect of speaker information on the prediction. The second model is further designed for phoneme-based TTS models and performs duration-aware pause insertion, predicting both RPs and punctuation-indicated pauses (PIPs) that are categorized by duration. The evaluation results show that our models improve the precision and recall of pause insertion and the rhythm of synthetic speech.
\end{abstract}

\begin{keywords} 
    multi-speaker TTS, pause insertion, phrase break prediction, phrasing, categorized pause insertion, BERT
\end{keywords}

\vspace{-2mm}
\section{Introduction} \label{sec:introduction}
\vspace{-2mm}
Human speakers usually insert silent pauses into speech to take a breath or show better expression. There are two main types of silent pause: respiratory pauses (RPs)~\cite{bailly12,klimnov17} and punctuation-indicated pauses (PIPs). The former is inserted at word transitions without punctuation to utter long sentences fluently, and the latter is inserted at punctuation marks following text descriptions. Pause insertion in text-to-speech (TTS) systems, also known as phrase break prediction and phrasing, is an essential part of making computers and robots speak as fluently as human speakers.

Since people always insert PIPs at punctuation marks, most previous works on phrasing have focused on the position prediction of RPs. Conventionally, linguistic information including lexical features (e.g., part-of-speech tags) and syntax features (e.g., distance from punctuation) is used for this task. Machine learning methods are used in phrasing models, such as decision tree algorithms~\cite{klimnov17,keri07,parlikar11,watts11,braunschweiler16,shi07,zhang16}, hidden Markov models~\cite{bell06,read07,chen15}, and conditional random fields~\cite{keri07,qian10}. Due to the development of natural language processing (NLP) and deep learning technologies, word representations have become the key linguistic feature. Moreover, recurrent neural networks (RNNs), especially bidirectional long short-term memory (BiLSTM)~\cite{graves05}, have become the mainstream models~\cite{klimnov17,vadapalli16,ding15,futamata21}. Bidirectional encoder representations from transformers (BERT)~\cite{devlin19}, one of the well-known pre-trained language models currently, also shows potential for this task. For example, Futamata et al. have introduced features from pre-trained BERT in Japanese phrase break prediction~\cite{futamata21}, and Abbas et al. have taken word-level BERT embeddings as the input of a conventional phrasing model~\cite{abbas22}.

However, the related works have not considered that various speakers have different styles of inserting RPs. Although some latent grammar and rules are shared among speakers, such differences can significantly reduce the accuracy of an RP insertion model when we train it using a multi-speaker corpus without speaker information. In fact, we always have to train the model with a multi-speaker corpus due to the sparse distribution of RPs. Furthermore, the speaker difference can also affect the insertion of PIPs, as well as the duration of the inserted pauses. For example, speakers do not always add a pause at every punctuation mark when reading text, depending on context and their habits.

Besides, in practice, most mainstream TTS models (e.g., FastSpeech~2~\cite{ren21} and Glow-TTS~\cite{kim20}) mainly use phonemes as input. They treat all silent pauses as one phoneme, which leads to the duration of all silent pauses in synthetic speech following the same distribution and not being sufficiently differentiated. From our observation in preliminary experiments, the longer the synthetic speech, the worse the rhythm due to the lack of well-differentiated pauses. Therefore, we expect the input of silent pause phonemes with duration information to enable phoneme-based TTS models to predict more accurate silent pauses.

This paper proposes two multi-speaker pause insertion models: the respiratory pause insertion (RPI) model and the categorized pause insertion (CPI) model. The architecture is based on BERT and BiLSTM. The RPI model is a phrasing model that considers the speaker difference by adding a speaker embedding to the hidden sequence output by BERT. It is used to show the improvement of pre-trained BERT and speaker information on RP prediction. With this RPI model as a basis, we further propose the CPI model for phoneme-based TTS models, in which both RPs and PIPs are categorized by duration, and their position and category are predicted. We present objective evaluations of phrasing accuracy and subjective evaluations of synthetic speech from text with automatically inserted RPs and PIPs. The results show that our models perform better than the baseline~\cite{klimnov17} and bring better rhythm to synthetic speech. Speech samples are available online\footnote{\url{https://ydqmkkx.github.io/pause-insertion/}}.

\vspace{-3mm}
\section{Dataset} \label{sec:dataset}
\vspace{-2mm}

We constructed the dataset from LibriTTS~\cite{zen19}, a multi-speaker English corpus derived from the audiobooks on the LibriVox website\footnote{\url{https://librivox.org}}. LibriTTS includes plenty of long-form sentences containing multiple silent pauses uttered by more than 2,000 speakers, and thus fits our purpose of evaluating the performance of multi-speaker pause prediction. 

First, we pre-processed the text. We converted all words to lowercase subwords. In the case of multiple consecutive punctuation marks, we kept only the first one for better cleaning and alignment. We used the Montreal Forced Aligner (MFA)~\cite{mcauliffe17} to align text and speech and obtain pause durations. Because MFA recognizes the silence at word transitions over 30~ms as silent pauses, we regarded silent pauses over 30~ms at punctuation marks as PIPs and those over 50~ms at word transitions without punctuation as RPs. We also kept the sentences that did not contain RPs in our dataset. 

Second, we categorized the silent pauses by duration through the Gaussian mixture model-based method in \cite{campione02}. The distribution of pause duration was fitted with several Gaussian distributions, after which the cut-off points of every two adjacent Gaussian distribution curves were found. For convenience, the whole hundred numbers nearest to the value on the horizontal axis of each cut-off point were used as the thresholds for categorization. We specified the pauses as brief ($<$ 300~ms), medium (300--700~ms), or long ($>$ 700~ms). 

Statistics of the dataset are shown in \Table{corpus_spec}, in which tokens consist of subwords and punctuation. Each token was annotated with the four types of labels shown in \Fig{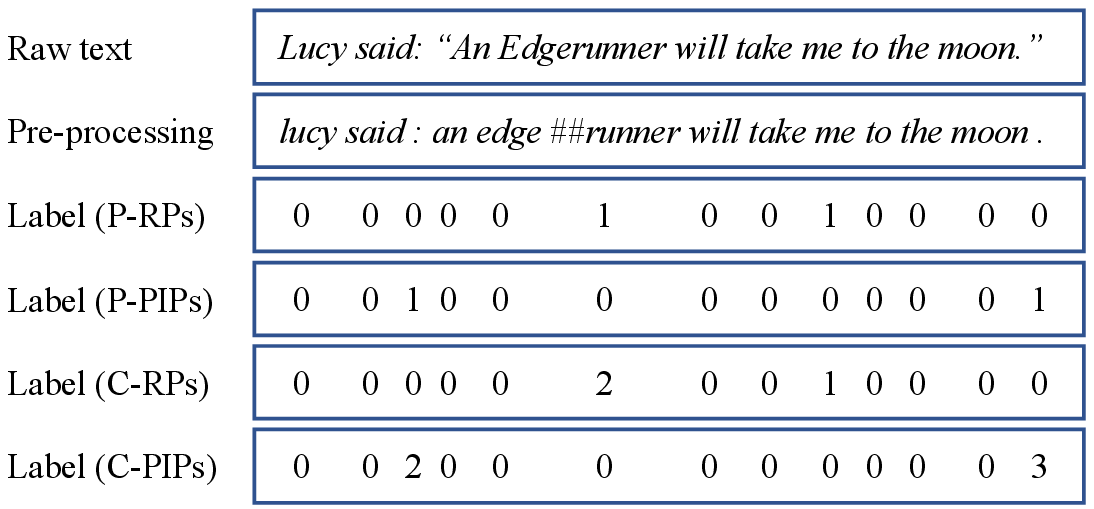}. Labels with the format P-x (x is the type of pauses) include information about the position of pauses. In P-RPs, the label ``1'' means an RP appears after its corresponding subword. We only labeled the last subword of a word as ``1''. The labels of C-x represent the categories of pauses. Category ``0'' means ``no pause'', and categories ``1'', ``2'', and ``3'' correspond to the brief, medium, and long pauses, respectively. 

\begin{figure}[t]
    \centering 
    \vspace{0mm}
    \includegraphics[width=\linewidth]{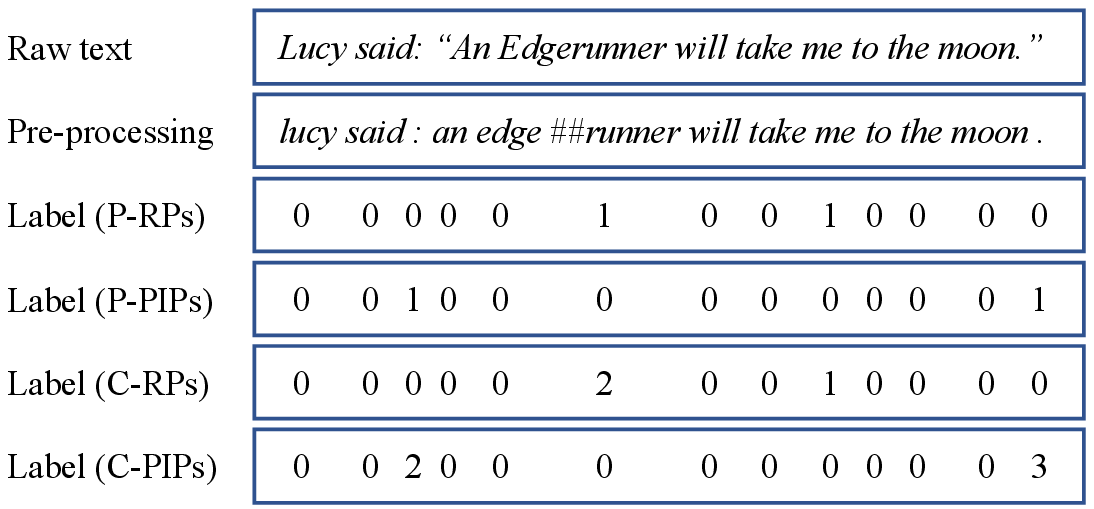} 
    \vspace{-4mm}
    \caption{An example of text pre-processing and label setting. \#\#: identifier of continuing subwords, P-x: position of x, C-x: category of x.}
    \label{fig:fig/pprp_example.eps}
    \vspace{-2mm}
\end{figure}

\begin{table}[t]
    \vspace{2mm}
    \centering
    \caption{Statistics of the dataset.}
    \vspace{-2mm}
    \begin{tabular}{l|rrr}
        \hline \hline
                    &     Training & Validation & Test    \\
        \hline
        Sentences   &   404,307 &  10,000    &  10,000 \\ 
       Tokens       & 9,098,772 & 225,700    & 225,638 \\
        Punctuation & 1,047,561 &  26,046    &  26,001 \\ 
        Speakers    &     2,305 &   2,154    &   2,162 \\ 
RPs (total)         &   170,168 &   4,083    &   4,297 \\
RPs (category 1)    &   136,306 &   3,287    &   3,450\\
RPs (category 2)    &    32,367 &     768    &     813\\
RPs (category 3)    &     1,495 &      28    &      34\\
PIPs (total)        &   861,544 &  21,329    &  21,354\\
PIPs (category 1)   &   399,559 &   9,898    &   9,979\\
PIPs (category 2)   &   325,327 &   8,060    &   7,953\\
PIPs (category 3)   &   136,658 &   3,371    &   3,422\\
        \hline \hline
    \end{tabular}
    \label{tab:corpus_spec}
\end{table}

\begin{figure*}[t]
    \centering 
    \vspace{0mm}
    \includegraphics[width=0.9\linewidth]{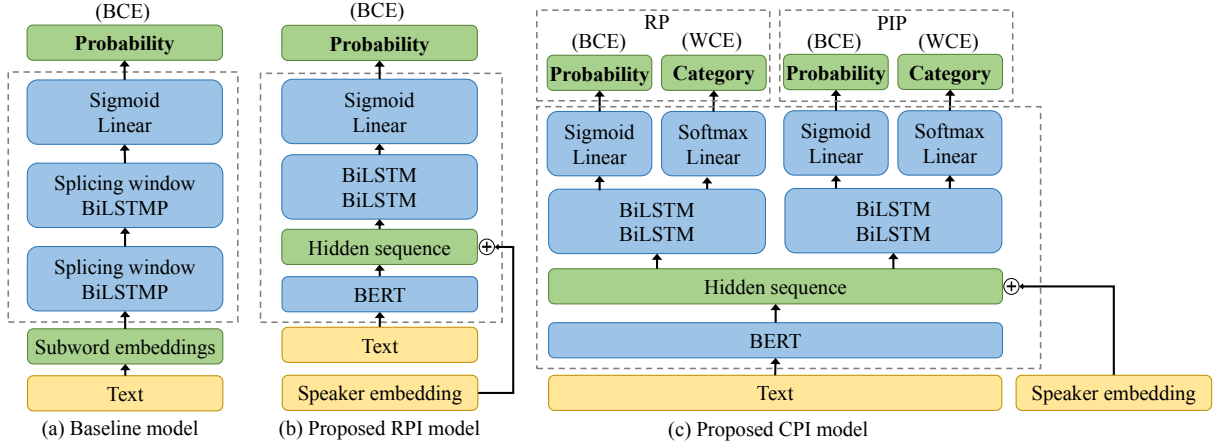} 
    \vspace{-3mm}
    \caption{Architectures of the baseline and proposed models.}
    \label{fig:fig/architecture.eps}
    \vspace{0mm}
\end{figure*}

\vspace{-3mm}
\section{Baseline Method} \label{sec:baseline}
\vspace{-2mm}

Klimkov et al.~\cite{klimnov17} provided the mainstream framework of a phrasing model based on English audiobooks, which focused on long-form reading TTS and was implemented as our baseline. The architecture of the baseline model is shown in \Fig{fig/architecture.eps}(a). The input word2vec embeddings are processed by BiLSTM projection with peephole connection (BiLSTMP)~\cite{sak14} layers and splicing windows. The splicing window stacks together seven frames before and after and feeds them to the next layer. The model outputs the probability of the occurrence of a pause after each token. 

In our implementation, we used subword embeddings as the input (with the same dimension as the word embeddings in the original paper) and replaced the softmax function with the sigmoid function for consistency with our proposed models.

\vspace{-3mm}
\section{Proposed Method} \label{sec:proposed}

\vspace{-2mm}
\subsection{RPI model} \label{subsec:RPI_model}
\vspace{-2mm}

The proposed RPI model (\Fig{fig/architecture.eps}(b)) is inspired by \cite{futamata21}. As with the baseline model, it predicts the position of RPs. BERT has demonstrated its unparalleled ability to extract textual features in various NLP tasks, and our task is well suited as its downstream task. However, it cannot receive and utilize speaker information very well by itself. Hence, we use the encoder-decoder structure and take BERT as the encoder. Two BiLSTM layers are then used to decode the information from BERT and speaker embeddings, which are initialized randomly and trained with the RPI model. The hidden sequence is the output of the last layer in BERT.

The main purpose of the RPI model is to quantify the improvements brought by the pre-trained BERT and speaker embeddings on phrasing.

\vspace{-2mm}
\subsection{CPI model} \label{subsec:CPI_model}
\vspace{-2mm}

As presented in \Fig{fig/pprp_example.eps} and stated in \Sec{dataset}, we categorized the silent pauses into three categories according to their duration. Then when training the TTS model, silent pauses in the dataset were represented as three marks. For example, ``sp'', which usually represents silent pauses in phonemes, was divided into ``sp1'', ``sp2'', and ``sp3'', corresponding to the three categories.

When converting a text sequence into a phoneme sequence during the inference of TTS models to insert PIPs, a common approach is to convert several specific types of punctuation into pause marks and ignore the other punctuation. This approach is a little crude because punctuation pauses also need to be predicted. To overcome this limitation, our CPI model (\Fig{fig/architecture.eps}(c)) predicts both RPs and PIPs (including their position and category) by using a multi-task learning framework. The CPI model utilizes two sets of BiLSTM layers to decode the hidden sequence from BERT and speaker embeddings that correspond to the predictions of RPs and PIPs, as the two pauses are different in the distributions of position and category. In our preliminary experiment, we used the same decoder to make both predictions. In that case, the prediction distribution of RPs was significantly influenced by that of PIPs, which led to poor results because there are many more PIPs than RPs.

As shown in \Fig{fig/architecture.eps}(c), in addition to the \textbf{Category}, the CPI model also outputs the \textbf{Probability} of pause occurrence, because taking the predictive probability of category 0 as that of pause occurrence is not very effective. Specifically, the model first predicts \textbf{Probability} that represents the occurrence of pauses and then outputs \textbf{Category} with the highest probability among categories 1--3.

\vspace{-2mm}
\section{Experimental Evaluations} \label{sec:experiment}
\vspace{-2mm}

\subsection{Experimental configurations}  \label{sec:experiment_conf}
\vspace{-2mm}

\begin{figure}[t]
    \centering 
    \vspace{0mm}
    \includegraphics[width=0.7\linewidth]{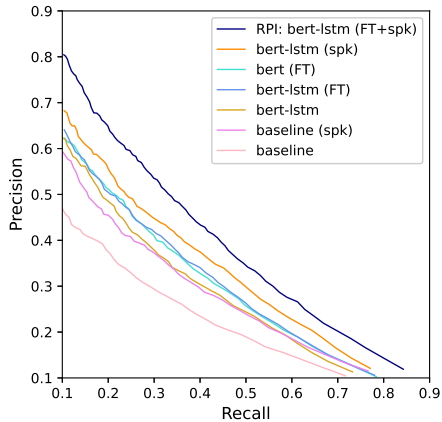} 
    \vspace{-3mm}
    \caption{Precision-recall curves on RP position prediction.}
    \label{fig:fig/recall_precision.eps}
    \vspace{-4mm}
\end{figure}

\begin{table}[t]
    \centering
    \vspace{2mm}
    \caption{Results of RP position prediction.}
    \vspace{-2mm}
    \begin{tabular}{l|rrr}
        \hline \hline
            & Precision & Recall & $\mathrm{F}_{0.5}$ \\
        \hline
    RPI: bert-lstm (FT+spk) & \textbf{0.569} &  \textbf{0.272} & \textbf{0.467} \\ 
  bert-lstm (spk) & 0.490 &  0.253 &  0.413\\
                   bert (FT) & 0.487 &  0.233 &  0.400\\
             bert-lstm (FT) & 0.467 &  0.246 &  0.396\\
          bert-lstm & 0.475 &  0.213 &  0.381\\
             baseline (spk) & 0.446 &  0.209 &  0.364\\
                     baseline & 0.393 &  0.187 &  0.322\\
        \hline \hline
    \end{tabular}
    \label{tab:obj_rp}
    \vspace{0mm}
\end{table}

For the baseline model, the hidden size and projection size for each BiLSTMP layer were 512 and 128, respectively. The dimension of subword embeddings was 300. For the proposed models, we configured the BERT as BERT$_{\rm BASE}$\footnote{\url{https://huggingface.co/bert-base-uncased}} and set the hidden size of each BiLSTM layer to 512. The dimensions of the hidden sequence and speaker embeddings were 768. The BERT$_{\rm BASE}$ was pre-trained on BookCorpus~\cite{zhu15}, which consists of unpublished books and English Wikipedia entries. We also used this corpus to pre-train the subword embeddings by a Continuous Bag-of-Words (CBOW) model~\cite{mikolov13}. When training the baseline and RPI models, we used the P-RPs label shown in \Fig{fig/pprp_example.eps}. Four labels, namely, P-RPs, C-RPs, P-PIPs, and C-PIPs, were the targets of the four outputs of the CPI model. In labels C-RPs and C-PIPs, category 0 represents no pause after the token, which can be viewed as a placeholder for the convenience of training. The loss function was binary cross-entropy (BCE) loss for \textbf{Probability} and weighted cross-entropy (WCE) loss for \textbf{Category}. The weights of categories 1--3 in WCE loss function equaled the total number of category 0 divided by their total number in the training set. As an exception, we set the weight of category 3 of RPs to 1.0 due to its sparseness.

During training, each mini-batch had 32 sentences. We used the Adam optimizer~\cite{kingma15} and set the initial value of the dynamic learning rate to $5\times10^{-5}$. The learning rate dropped by 0.2 times when the model's performance on the validation set did not improve within 5000 iterations. We took 200,000 as the maximum iterations (about 16 epochs). The models that performed best on the validation set during these iterations were saved to make comparisons. We utilized the unbalanced F-score described in \cite{klimnov17} for threshold selection:
\begin{align}
    \mathrm{F}_{\beta} &= \pt{1 + \beta^{2}} \frac{\mathrm{Precision} \times \mathrm{Recall}}{\beta^{2} \times \mathrm{Precision} + \mathrm{Recall}} 
    \label{eq:threshold}
\end{align}
The maximal value of $\mathrm{F}_{\beta}$ that the model could achieve was taken as a criterion for evaluation. It is preferable to skip RPs rather than to insert them in inappropriate places~\cite{klimnov17}, so we took 0.5 as the value of $\beta$. For PIPs, we focused more on the recall and used $\mathrm{F}_{2}$. From our point of view, the absence of some critical PIPs makes for long utterances with bad rhythm.

\vspace{-2mm}
\subsection{Objective evaluations} \label{subsec:obj_eval}

\subsubsection{RPI model} \label{subsubsec:RPI}

This objective evaluation was designed to explore and quantify the improvements of the pre-trained BERT, fine-tuning of BERT, and introducing speaker information on phrasing models. We made comparisons on the position prediction of RPs. To examine the effect of adding speaker embeddings, we also added the speaker embedding after the first splicing window layer of the baseline model to take it as an encoder-decoder model, which we denote as ``baseline (spk)''. The whole structure of the proposed RPI model is denoted as ``bert-lstm (FT+spk)''. To determine the performance of the models under different thresholds, we used precision-recall curves to show the results and only kept the points in the curves whose both precision and recall values were above 0.1. When calculating the precision and recall values for the multiple subwords from one word, we counted only the last subword with the same rules as setting labels. From the results shown in \Fig{fig/recall_precision.eps} and \Table{obj_rp}, we can find:
\begin{itemize}[leftmargin=3mm,topsep=0mm,itemsep=-1mm]
    \item The BERT-based models performed better than the baseline models, which suggests that the pre-trained BERT can provide more efficient textual features than conventional word2vec models. 
    \item There were only slight differences between \textbf{bert-lstm (FT)} and \textbf{bert (FT)}, which means that without speaker embedding, BERT alone is capable of predicting the position of RPs after fine-tuning. 
    \item When adding speaker embeddings, the curve of \textbf{baseline (spk)} was close to that of \textbf{bert-lstm}, and the performance of \textbf{bert-lstm (spk)} even exceeded \textbf{bert-lstm (FT)}. These enhancements demonstrate the validity and generalizability of using speaker embeddings and show that different speakers have different styles for inserting RPs. For a large multi-speaker dataset, such differences deserve attention and can significantly affect the accuracy of RP prediction models.
    \item \textbf{bert-lstm (FT)} obtained a bit higher precisions than \textbf{bert-lstm} for the same recalls, which means that the fine-tuning of BERT leads to better performance but not by much without speaker embedding. However, the BERT-based model with speaker embeddings got a big boost from fine-tuning, which shows the need to apply them together. 
\end{itemize}

\vspace{-2mm}
\subsubsection{CPI model} \label{subsubsec:CPI}

\Table{obj_pp} and \Table{confmat} show the results of the proposed CPI model for position prediction and category prediction of both pauses, respectively. It is clear that the model retained the ability to predict the position of RPs, and achieved a recall value over 0.99 for the relatively simple task of position prediction of PIPs. In category prediction, the accuracy of category 2 was lower than that of the other two categories, which suggests that there is still some room for improvement in the choice of thresholds for categorization.

\begin{table}[t]
    \centering
    \caption{Results of position prediction of CPI.}
    \vspace{-2mm}
    \begin{tabular}{l|rrr}
        \hline \hline
            & Precision & Recall & $\mathrm{F}_{\beta}$ \\
        \hline
        RPs &     0.575 &  0.261 & $\mathrm{F}_{0.5}=0.463$ \\ 
       PIPs &     0.848 &  0.996 & $\mathrm{F}_{2}=0.962$ \\
        \hline \hline
    \end{tabular}
    \label{tab:obj_pp}
\end{table}
\begin{table}[t]
  \centering
  \caption{Confusion matrix of category prediction of RPs and PIPs.}\label{tab:confmat}
  \vspace{-2mm}
  \subfloat[Prediction of RPs]{%
    \begin{tabular}{c|rrr}
        \hline \hline 
        & \multicolumn{3}{|c}{Prediction}\\
        \hline 
        Label & 1 & 2 & 3 \\
        \hline
        1 & \textbf{2,565} & 885 & 0  \\
        2 & 300 & \textbf{513} & 0 \\
        3 & 14 & 20 & \textbf{0} \\
        \hline \hline
    \end{tabular}
  }\hspace{0.5cm}
  \subfloat[Prediction of PIPs]{%
    \begin{tabular}{c|rrr}
        \hline \hline 
        & \multicolumn{3}{|c}{Prediction}\\
        \hline 
        Label & 1 & 2 & 3 \\
        \hline
        1 & \textbf{6,155} & 1,766 & 2,058  \\
        2 & 2,258 & \textbf{3,186} & 2,509 \\
        3 & 335 & 352 & \textbf{2,735} \\
        \hline \hline
    \end{tabular}
  }
  \vspace{-4mm}
\end{table}
\begin{table}[t]
    \centering
    \caption{Subjective performance of pause insertion models. *: the model takes unmatched speaker embedding as input.}
    \vspace{-2mm}
    \label{tab:abtest}
    \begin{tabular}{l|c|c|c}
        \hline \hline
        Method A & Score & Method B & p-value \\
        \hline
        RPI & \textbf{0.560 vs. 0.440} & FastSpeech2 & $< 0.005$ \\
        RPI & 0.537 vs. 0.463          & Baseline    & $< 0.1$  \\
        CPI & \textbf{0.557 vs. 0.443} & Baseline    & $< 0.01$ \\
        RPI & {0.488 vs. 0.512}        & CPI (Position) & 0.60 \\
        RPI & \textbf{0.460 vs. 0.540} & CPI  & $< 0.05$ \\
        RPI & 0.510 vs. 0.490          & RPI* & 0.62     \\
        CPI & \textbf{0.550 vs. 0.450} & CPI* & $< 0.05$ \\
        \hline \hline
    \end{tabular}
\end{table}

\vspace{-2mm}
\subsection{Subjective evaluations} \label{subsec:sbj_eval}
\vspace{-2mm}

To explore the performance of our proposed models in multi-speaker TTS, especially to show the improvement of inputting categorized pause phonemes, we performed AB preference tests using FastSpeech 2 as our TTS model with HiFi-GAN~\cite{kong20} as the vocoder. For RPI and CPI models, the thresholds of position predictions were as stated in Section \ref{sec:experiment_conf}. We used ``clean'' subsets in the LibriTTS corpus to train two TTS models with silent pause phonemes, one with non-categorized pauses (``sp''), and one with categorized pauses (``sp1'', ``sp2'', and ``sp3''). In total, we selected 16 speakers (eight men, eight women) with good synthetic sound quality as the test speaker set in our evaluations. 

To ensure the synthetic speech in our test contained sufficient RPs and PIPs, we first selected long-form sentences with a total number of words and punctuation marks between 50 and 60 from the test set. After this, 123 sentences were left, each of them including more than one RPs. These sentences were then used in TTS synthesis, where one sentence corresponds to several speakers. To prevent the sound quality from affecting the tests, we removed any synthetic speech with bad quality and kept 277 text-speaker pairs. 

To explore whether the listener could perceive if a pause insertion style did not match the speaker, for each text-speaker pair, we utilized another speaker embedding from the test speaker set to do the pause insertion and made the gap as large as possible (denoted by * in \Table{abtest}). In addition, to show that the improvement in CPI relative to RPI came more from inputting categorized pause phonemes rather than from the extra position prediction of PIPs, we made CPI only predict uncategorized pauses and denoted it as ``CPI (Position)''. We asked native listeners from Amazon Mechanical Turk to participate in the tests. Every test was completed by 30 listeners, each of whom listened to ten pairs of synthetic speech and was asked to choose the one with better rhythm. 

We set seven tests. As shown in \Table{abtest}, we can find: 
\begin{itemize}[leftmargin=3mm,topsep=-1mm,itemsep=-1mm] 
    \item Listeners perceived the difference between \textbf{RPI} and \textbf{Baseline} to be insignificant and were insensitive to the position variance of RPs between \textbf{RPI} and \textbf{RPI*}. We believe this is due to the sparse distribution and short duration of RPs. People only became aware of the absence of RP when listening to a long sentence without a pause, where the position of RPs could not attract enough attention. 
    \item The slight lead of \textbf{RPI} over \textbf{Baseline} could also be due to the fact that RPI inserts more RPs with the same requirement of $\mathrm{F}_{\beta}$ value. 
    \item Listeners were more likely to recognize that \textbf{CPI} performed better than \textbf{Baseline} and \textbf{RPI}. Coupled with the inability of the listeners to distinguish significantly between \textbf{RPI} and \textbf{CPI (Position)}, we can conclude that inputting categorized pause phonemes to phoneme-based TTS models makes the rhythm of synthetic speech better. 
    \item According to the \textbf{CPI} and \textbf{CPI*} pair, listeners were sensitive to the difference arising from inserting different categories of pauses. Since the duration variation of pauses is present in both RPs and PIPs, using other speaker embeddings brings clear differences.
\end{itemize}

\vspace{-2mm}
\section{Conclusion} \label{sec:conclusion}
\vspace{-2mm}
In this paper, we proposed a respiratory pause insertion model and a categorized pause insertion model. The results of objective evaluations demonstrated that the speaker information can bring a large improvement to phrasing models trained with a multi-speaker dataset. The results of subjective evaluations showed that by inserting categorized pauses, the synthetic speech had better rhythm and was more consistent with the features of the speaker. In future work, we plan to explore the effectiveness of incorporating speaker embedding into the text-processing model of the TTS system for other similar tasks.

\bibliographystyle{bib/IEEEtran}
\bibliography{bib/refs}

\end{document}